\documentclass[preprint,12pt]{elsarticle}

\usepackage{amssymb}
\usepackage{axodraw}

\newcounter{bla}

\journal{Computer Physics Communications}

\begin{document}

\begin{frontmatter}



\title{Monte Carlo event generator for black hole production and decay
in proton-proton collisions\\ 
-- QBH version 1.03}


\author{Douglas M. Gingrich}

\ead{gingrich@ualberta.ca}
\address{Centre for Particle Physics, Department of Physics, University
of Alberta,\\ Edmonton,AB T6G 2G7 Canada}
\address{TRIUMF, Vancouver, BC V6T 2A3 Canada}

\begin{abstract}
We describe the Monte Carlo event generator for black hole production
and decay in proton-proton collisions -- QBH version 1.03. 
The generator implements a model for quantum black hole production and
decay based on the conservation of local gauge symmetries and
democratic decays.
The code in written entirely in C++ and interfaces to the PYTHIA~8
Monte Carlo code for fragmentation and decays.
\end{abstract}


\end{frontmatter}



{\bf PROGRAM SUMMARY/NEW VERSION PROGRAM SUMMARY}

\begin{small}
\noindent
{\em Manuscript Title:} Monte Carlo event generator for black hole
production and decay in proton-proton collisions\\
{\em Authors:} Douglas M. Gingrich\\
{\em Program Title:} QBH\\
{\em Journal Reference:}                                      \\
{\em Catalogue identifier:}                                   \\
{\em Licensing provisions:} none\\
{\em Programming language:} C++\\
{\em Computer:} x86\\
{\em Operating system:} Scientific Linux, Mac OS X\\
{\em RAM:} 1 GB
{\em Classification:} 11.6\\
{\em External routines/libraries:} PYTHIA 8 and LHAPDF\\
{\em Nature of problem:} Simulate black hole production and decay in
proton-proton collisions\\
\\
{\em Solution method:} Monte Carlo simulation using importance sampling\\
\\
   \\
{\em Running time:} Eight events per second\\
   \\

\end{small}

\section{Description of Generator}

Black holes are produced in proton-proton collisions and decay to
two-particle final states~\cite{Gingrich09b}. 
Electric charge, QCD colour, and spin are conserved, while baryon and
lepton number need not be conserved. 
PYTHIA~8 is used for the initial- and final-state radiation, multiple
interactions, final state parton showers, fragmentation, and decay
processes~\cite{pythia8,pythia64}. 
The QBH generator interfaces to PYTHIA~8 using the Les Houches
accord (LHA)~\cite{LHA}.

The first black hole generators were TRUENOIR written by Dimopoulos
and Landsberg\footnote{http://hep.brown.edu/users/Greg/TrueNoir/}
\cite{Dimopoulos01},
and an unnamed generator used in the first black hole study specific
to ATLAS by Tanaka \textit{et al.}~\cite{Tanaka}. 
These generators were made quick and efficient for specific studies by
averaging over some dynamical quantities, rather than generating them
according to probabilistic distributions.
The first two true Monte Carlo event generators for black hole
production and decay were CHARYBDIS~\cite{Harris03a} and
CATFISH~\cite{Cavaglia06}.    
A comparison between CHARYBDIS and CATFISH can be found in
Ref.~\cite{Gingrich06b}. 

The most recent generators CATFISH~\cite{Cavaglia06},
Charybdis~2~\cite{Frost09}, and BlackMax~\cite{Dai09} all include the
facility to generate black holes decaying to two-particle final states. 
It is possible to generate two-particle final states with CATFISH but
there are some caveats. 
One sets the final $N$-body decay to two particles and raises the
black hole mass threshold at which Hawking evaporation ends
$Q_\mathrm{min}$ to be above the minium black hole mass and Planck
scale.  
If $Q_\mathrm{min}$ is set too low, one will occasionally produce black
holes above $Q_\mathrm{min}$ resulting in more than two particles in
the final state.
If $Q_\mathrm{min}$ is set too high, one will occasionally produce black 
holes with high mass that only decay to two-particle final states rather
than the anticipated Hawking evaporation.
BlackMax implements the suggestions of Meade and Randall
verbatim~\cite{Meade08}. 
The decay species are chosen according to thermal greybody-modified
distributions, which is perhaps not applicable to black holes near the
Planck scale~\cite{Gingrich09b}. 
Charybdis~2 allows the user to set parameters so that effectively a
black hole decaying to two-particle final states is generated.
CATFISH and Charybdis~2 do not take colour explicitly into account when
determining the two-particle final state.
In all three cases, the generators do not allow for the ability to
generate a particular black hole state, which is a representation of
$SU(2)$ colour, charge, and spin.  
They thus act as inclusive black hole generators and sum over all
possible partons giving rise to black holes. 
In addition, these generators do not allow the ability to easily
generate jet-lepton or jet-boson topologies.   
Although baryon number can be violation by these generators when run in
standalone mode, the partons in the output record can not be hadronized
by programs that are typically used to provide the hardonization,
fragmentation, and decay process.
QBH can be used to select the specific black hole state to
generate and includes all possible decay modes properly weighted by
conserving $SU(3)_c$ and $U(1)_\mathrm{em}$.
In the following, we refer to these different black holes states as
quantum black holes, with the understanding that we are considering a
model which is an extrapolation from the classical regime~\cite{Gingrich09b}. 
All the QBH code, including PYTHIA~8, is written in C++.

\section{How to use the Generator}

The generator consists of four C++ classes, the main class
\texttt{QuantumBlackHole}, a derived \texttt{LHAup} user class
\texttt{dLHAup}, and two utility classes \texttt{Decay} and \texttt{Random}.  
The user normally only needs to interface to the \texttt{QuantumBlackHole}
class, and only for initialization. 

The package is typically used as two files:
the header file \texttt{qbh.h} containing class definitions and the file
\texttt{qbh.cc} containing the bodies of the member functions.

The user must supply his own main program to initialize the package and
generate events.
The procedure is very similar to writing a standalone PYTHIA~8 main
program. 
Above the main program, the following lines must appear in this order.

\bigskip
\begin{ttfamily}
\noindent
\#include \symbol{34}Pythia.h\symbol{34};\\
using namespace Pythia8;\\
\#include \symbol{34}qbh.h\symbol{34};\\
using namespace QBH;\\
\end{ttfamily}

Near the top of the main program, the three objects
\texttt{QuantumBlackHole}, \texttt{dLHAup},  and \texttt{Pythia} need to
be allocated.  
The following is an example of creating them dynamically.

\bigskip
\begin{ttfamily}
\noindent
Pythia* pythia = new Pythia();\\
dLHAup* lhaPtr = new dLHAup();\\
QuantumBlackHole* qbh = new QuantumBlackHole(pythia,true);\\
\end{ttfamily}

\noindent
The second argument to the \texttt{QuantumBlackHole} constructor needs
to be \texttt{true} for initialization.
After initialization, the \texttt{QuantumBlackHole} object may be
deleted. 

The user is free to further initialize any QBH or Pythia variables
before the final initialization of the generator.
The final initialization of the generator is made by passing a pointer
to the derived class \texttt{dLHAup} to PYTHIA as follows.

\bigskip
\begin{ttfamily}
\noindent
bool status = pythia->init(lhaPtr);\\
\end{ttfamily}

\noindent
After this, the main program looks like any other PYTHIA main program.
The user is free to generate events and study these events using all the
machinery of the PYTHIA classes.

Les Houches accord event files~\cite{LHALEF} may be written using the
derived class \texttt{dLHAup} member functions \texttt{openLHEF},
\texttt{initLHEF}, \texttt{eventLHEF}, and \texttt{closeLHEF}. 

\subsection{User Initialization}

The QBH package initializes automatically with a sensible set of
parameters.
But typically the user will want to change a few parameters to set up
the run of interest.
The parameters are available as static global variables of the
\texttt{QuantumBlackHole} class.
The first time the \texttt{QuantumBlackHole} constructor is call with a
boolean argument that is true, a more detailed initialization occurs. 
Any further instantiations of the class, with or without arguments, does
no further initialization.
However, the user is free to initialize any of the parameters using a
collection of ``set'' methods.

The constructor sets the lower bound on the quantum black hole mass to
the Planck scale and the upper bound on the mass to three times the
Planck scale or the proton-proton centre of mass energy, whichever is
less.  
The constructor also initializes the LHAPDF parton distribution
functions, which are the default for QBH, and calculates the branching
ratios. 
Lower bounds on the cross section using the calculations of Yoshino
and Rychkov~\cite{Yoshino05a} may be used.
In this case, the constructor reads in the appropriate data file.

Normally the user uses the ``set'' methods to set the quantum black hole
state, Planck scale, number of dimensions, etc.
For example, the quantum black hole state is set using the member
functions \texttt{setQstate}, \texttt{setIstate}, and the Planck scale
is set using \texttt{setMplanck}.   
Table~\ref{tab1} shows a complete set of ``set'' methods available to
the user.

\begin{table}[p]
\caption{\texttt{QuantumBlackHole} class public member functions for setting
parameters. 
The default values are shown as (D=...). 
\label{tab1}}
\begin{center}
\begin{tabular}{ll}
\texttt{setQscale(bool)}:   & (D=true) definition of QCD scale for PDFs.\\ 
                            & = false: quantum black hole mass,\\
                            & = true:  inverse gravitational radius.\\
\texttt{setYRform(bool)}:   & (D=false) use Yoshino-Rychkov
                              factors~\cite{Yoshino05a}.\\  
\texttt{setTrap(bool)}:     & (D=false) use Yoshino-Rychkov trapped
                              surface calculation~\cite{Yoshino05a}.\\  
\texttt{setRS1(bool)}:      & (D=false) RS or ADD black hole.\\
                            & = false: ADD black hole,\\
                            & = true:  Randall-Sundrum type-1 black hole.\\
\texttt{setSM(bool)}:       & (D=true) conserve global symmetries.\\
\texttt{setMajorana(bool)}: & (D=false) neutrinos are Majorana or Dirac.\\
                            & = false: Dirac,\\
                            & = true: Majorana.\\
\texttt{setChiral(bool)}:   & (D=true) handedness of neutrinos.\\
                            & = false: neutrinos are left-handed only,\\
                            & = true: neutrinos left-handed and right-handed.\\
\texttt{setHiggs(bool)}:    & (D=true) include Higgs as allowed particle.\\
\texttt{setGraviton(bool)}: & (D=true) include graviton as allowed particle.\\
\texttt{setTotdim(int)}:    & (D=10) total number of spacetime
                              dimensions.\\
                            & = $5-11$ allowed.\\
\texttt{setPlanckdef(int)}: & (D=3) definition of Planck
                              scale~\cite{Gingrich06a}.\\ 
                            & = 1: Randall-Sundrum,\\
                            & = 2: Dimopoulos-Landsberg definition,\\
                            & = 3: PDG definition,\\
                            & = else: Giddings-Thomas definition.\\
\texttt{setLHAglue(int)}:   & (D=10042) LHA glue code~\cite{LHAPDF}.\\
                            & = 10042: LHA cteq6ll.LHpdf,\\
                            & = 19050: LHA cteq5m.LHgrid,\\
                            & = 29041: LHA MRST98l.LHgrid,\\
                            & = else: CTEQ5L (internal PYTHIA~8).\\
\texttt{setQstate(int)}:    & (D=9) three times electric charge.\\ 
                            & = $4, 3, 2, 1, 0, -1, -2, -3, -4$ allowed,\\ 
                            & = 9 used for all partons.\\ 
\texttt{setIstate(int)}:    & (D=3) initial state.
                             = 0: $q$-$q$,
                             = 1: $q$-$g$,
                             = 2: $g$-$g$,
                             = 3: all.\\ 
\texttt{setMplanck(double)}: & (D=1000.0 GeV) fundamental Planck scale.\\
\texttt{setMinmass(double)}: & (D=1000.0 GeV) minimum quantum black
                                hole mass.\\ 
\texttt{setMaxmass(double)}: & (D=14000.0 GeV) maximum quantum black
                               hole mass.\\ 
\texttt{setEcm(double)}:     & (D=14000.0 GeV) proton-proton centre of
                               mass energy.\\ 
\end{tabular}
\end{center}
\end{table}

The method \texttt{setMaxmass}, along with \texttt{setMinmass}, allow
the user to set a range of black hole masses.
This could be useful for restricting the mass range for quantum black
holes and thus cutting out semiclassical black hole, sampling the
differential cross section by using narrow mass bins, or for user
debugging purposes.  
If the user sets the maximum black hole mass above the centre of mass
energy, set by method \texttt{setEcm}, the program at initialization will
reset the maximum black hole mass to be the centre of mass energy.

There are a corresponding set of methods which return the values of the
parameters: 
\texttt{ecm()},
\texttt{mplanck()},
\texttt{minmass()}, 
\texttt{maxmass()},
\texttt{qstate()},
\texttt{istate()},
\texttt{totdim()},
\texttt{planckdef()},
\texttt{lhaglue()},
\texttt{qscale()},
\texttt{yrform()},
\texttt{trap()},
\texttt{poisson()},
\texttt{RS1()},
\texttt{sm()},
\texttt{chiral()},
\texttt{majorana()},
\texttt{higgs()}, and
\texttt{graviton()}.

It is necessary to define some of the parameter here so their meaning is
clear to the user.  
A detailed description of the model can be found in
Ref.~\cite{Gingrich09b}.
The mass scale used by the Randall-Sundrum black hole
(\texttt{setPlanckdef(1)}) is defined as 

\begin{equation}
\tilde{M} \equiv \frac{m_1}{x_1 c^{2/3}}\, ,
\end{equation}

\noindent
where $m_1$ is the first KK graviton mass, $x_1 = 3.83$, and $c \equiv
k/M_\mathrm{Pl}$, where $k$ is the ADS curvature and $M_\mathrm{Pl}$ is
the reduced Planck scale.
In ADD, the fundamental Planck scale in the Dimopoulos-Landsberg, PDG,
and Giddings-Thomas definitions are, respectively,

\begin{equation}
M_\mathrm{DL}^{D-2} \equiv \frac{(2\pi)^{D-4}}{8\pi G_D},\quad\quad
M_\mathrm{D}^{D-2}  \equiv \frac{1}{G_D},\quad\quad
M_\mathrm{GT}^{D-2} \equiv \frac{(2\pi)^{D-4}}{4\pi G_D}\, ,
\end{equation}

\noindent
where $G_D$ is the $D$-dimensional Newton constant in the Myers-Perry metric.

\subsection{User Process Run Information}

The PYTHIA initialization call \texttt{pythia->init(lhaPtr)} calls the
virtual function \texttt{setInit()} of the derived class \texttt{dLHAup}
to initialize the physics process in the \texttt{QuantumBlackHole}
class. 
Virtual function \texttt{setInit()} calls
\texttt{QuantumBlackHole::init()} to calculated the total proton-proton
cross section, and then \texttt{QuantumBlackHole::banner(lhaPrt)} to
write a banner to the output.  
The banner lists all the parameters, the beam, and quantum black hole
configuration, as well as, the cross section and estimated statistical
error.
Some simple consistency checks are made, and the user may be warned or
asked to supply different parameters.

The internal processing of the run information is described next.
Although the user normally does not touch this part of the class, it is
informative to know how the Les Houches Accord is implemented.
The generator assumes both beams are protons to use the symmetry of the
collision to gain some efficiency.
The LHA weighting strategy is declared by setting the master switch
\texttt{IDWTUP} to $+1$. 
The generator supplies positive weighted events, and PYTHIA returns
events with weight 1 by using an accept or reject algorithm.
Only one physics process is declared in the interface to PYTHIA.
The selection of which quantum black hole to generate is determined by
QBH according to the user settings.
The user may select any of the 14 quantum black holes states for
generation, or a mixture of all the states weighted by their relative
cross section is possible (see Table~\ref{tab2}).

\renewcommand{\arraystretch}{1.2}
\begin{table}[htb]
\caption{\texttt{Qstate} and \texttt{Istate} codes for the 14 quantum
black hole states. 
\label{tab2}}
\begin{center}
\begin{tabular}{lcc}\hline
State & Qstate & Istate\\\hline
$QBH^{4/3}_{uu}$              &  4 & 0\\
$QBH^1_{u\bar{d}}$            &  3 & 0\\
$QBH^{2/3}_{ug}$              &  2 & 1\\
$QBH^{2/3}_{\bar{d}\bar{d}}$  &  2 & 0\\
$QBH^{1/3}_{\bar{d}g}$        &  1 & 1\\
$QBH^{1/3}_{ud}$              &  1 & 0\\
$QBH^0_{q\bar{q}}$            &  0 & 0\\
$QBH^0_{gg}$                  &  0 & 2\\
$QBH^{-1/3}_{\bar{u}\bar{d}}$ & -1 & 0\\
$QBH^{-1/3}_{dg}$             & -1 & 1\\
$QBH^{-2/3}_{dd}$             & -2 & 0\\
$QBH^{-2/3}_{\bar{u}g}$       & -2 & 1\\
$QBH^{-1}_{d\bar{u}}$         & -3 & 0\\
$QBH^{-4/3}_{\bar{u}\bar{u}}$ & -4 & 0\\
all & N/A & 3\\
\hline
\end{tabular}
\end{center}
\end{table}

\subsection{User Process Event Information}

The user generates an event by calling \texttt{pythia->next()} which
calls the virtual function \texttt{setEvent()} of the derived class
\texttt{dLHAup} to generate one event with the \texttt{QuantumBlackHole}
class. 
Virtual function \texttt{setEvent()} calls
\texttt{QuantumBlackHole::event()} to generate an event.

The differential cross section is integrated and sampled using
importance sampling. 
The integration and importance sampling is very fast since the cross
section is only a function of two variables and transformations are
applied to flatten the function that is sampled.

Sometimes the sampled cross section is vanishingly small.
This can occur near the edges of the parameters space or if the
trapped surface cross section is being calculated.
In these cases, the partons are both set to up-quarks and
\texttt{setEvent()} still returns \texttt{true}.
If \texttt{false} was returned, the cross section calculated by PYTHIA
would not be the same as that calculated by QBH.

\section{\texttt{QuantumBlackHole} Class Member Functions}

\begin{figure}[htb]
\begin{center}
\includegraphics[width=\columnwidth]{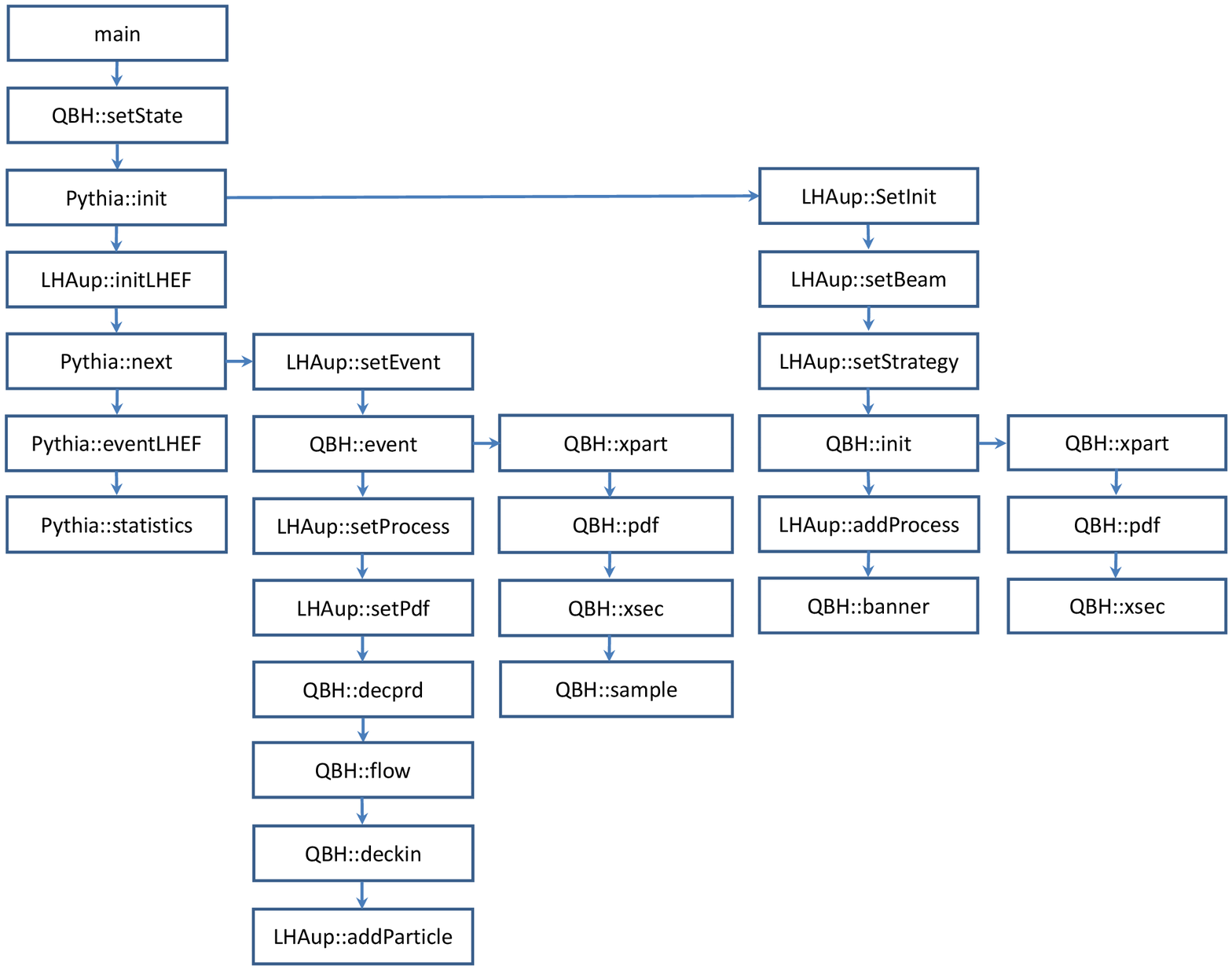}
\caption{Program flow through the member functions.
\texttt{xpart}, \texttt{pdf}, \texttt{xsec}, and \texttt{sample} are
private member functions. 
\label{fig9}}
\end{center}
\end{figure}

Figure~\ref{fig9} shows the program flow through the
\texttt{QuantumBlackHole} member functions.
Each public member function is described next.

\bigskip
\noindent
bool dLHAup::setInit(void)

\begin{tabular}{ll}
Purpose:          & Les Houches initialization virtual function.\\
Objects declared  & QuantumBlackHole.\\
Functions called: & LHAup::setBeamA, setBeamB, setStrategy, addProcess,\\
                  & QuantumBlackHole::init, banner.\\
Return:           & always true.\\
\end{tabular}
\bigskip

\noindent
\texttt{setInit} sets the beams characteristics, declares the generation
strategy, and defines a single physics process.
It also calculates the total proton-proton cross section and prints a
program banner on the standard output.

\bigskip
\noindent
bool dLHAup::setEvent(int idProcess)

\begin{tabular}{ll}
Purpose:          & Les Houches event virtual function.\\
Objects declared  & QuantumBlackHole.\\
Functions called: & LHAup::setProcess, setPdf, addParticle,\\
                  & QuantumBlackHole::event, decprd, flow, deckin.\\
Return:           & always true.\\ 
\end{tabular}
\bigskip

\noindent
\texttt{setEvent} sets the weight for each event, sets some PDF information,
and adds the beam and generated hard scattering particles to the event
record. 
It also adds the quantum black hole state to the event record for
documentation and process identification purposes.
It generates the event by calculating a weight for the event and
determines random $Q$, $x$ and beam particle IDs.
It determines the quantum black hole decay products, connects the colour
flow, and determines the mothers of the partons.  
The kinematics of the decay are calculated.

\bigskip
\noindent
QuantumBlackHole:QuantumBlackHole(Pythia*,bool initialize)

\begin{tabular}{ll}
Purpose:          & QuantumBlackHole constructor.\\
Functions called: & Many set methods.\\
Return:           & All user parameters, initializes PDFs, and
                    calculates branching\\
                  & ratios.\\ 
\end{tabular}
\bigskip

\noindent
The constructor sets information about the quantum black hole and extra 
dimensions.
These parameters should be fixed for the entire execution of the
program. 

\bigskip
\noindent
void QuantumBlackHole::init(double\& dxsec)

\begin{tabular}{ll}
Purpose:          & Quantum black differential cross section calculation.\\
Functions called: & QuantumBlackHole::xpart, pdf, xsec.\\
Return:           & proton cross section (pb) at random $x$ values.\\
\end{tabular}
\bigskip

\noindent
\texttt{init} calculates the parton differential cross section at a
random $x_\mathrm{min}$.
It then determines two random $x$ values and stores the PDFs. 
It calculates the proton-proton differential cross section at these
$x$ values. 

\bigskip
\noindent
void QuantumBlackHole::banner(dLHAup* inf)

\begin{tabular}{ll}
Purpose:          & Output initialization information.\\
Functions called: & none.\\
Return:           & none.\\
\end{tabular}
\bigskip

\noindent
\texttt{banner} outputs initialization information in a banner.

\bigskip
\noindent
bool QuantumBlackHole::event(double\& wxsec,double\& Q,double* xx,int* idn)

\begin{tabular}{ll}
Purpose:          & Quantum black hole event generation.\\
Functions called: & QuantumBlackHole::xpart, pdf, xsec, sample.\\ 
Return:           & always true.\\ 
\end{tabular}
\bigskip

\noindent
\texttt{event} calculates the parton cross section at a random
$x_\mathrm{min}$ value and returns the resulting differential cross
section and $Q$ value. 
It then determines two $x$ values and uses these to store the PDFs.
The proton-proton cross section is sampled to determine an event weight
and the two partons that caused the process are returned.

\bigskip
\noindent
void QuantumBlackHole::decprd(int *idn)

\begin{tabular}{ll}
Purpose:          & Determine decay products identifiers.\\
Functions called: & ParticleDataTable::chargeType. \\
Return:           & PDG ID codes of decay particles.\\
\end{tabular}
\bigskip

\noindent
\texttt{decprd} determines the decay products based on predetermined branching
ratios. 

\bigskip
\noindent
void QuantumBlackHole::flow(int *idn,int moth[5][2], int icol[5][2])

\begin{tabular}{ll}
Purpose:          & Determine the mothers and colour connections.\\
Functions called: & none.\\
Return:           & Mother and colour connections.\\
\end{tabular}
\bigskip

\begin{figure}[htb]
\begin{center}
\begin{picture}(80,80)(-40,-40)
\SetWidth{0.75}
\ArrowLine(-25,-25)(0,0)
\ArrowLine(0,0)(25,-25)
\ArrowLine(-25,25)(0,0)
\ArrowLine(0,0)(25,25)
\SetColor{Red}
\ArrowLine(-25,-35)(0,-10)
\ArrowLine(0,-10)(25,-35)
\SetColor{Blue}
\ArrowLine(-25,35)(0,10)
\ArrowLine(0,10)(25,35)
\Text(-30,-30)[rb]{$q$}
\Text(-30, 30)[rt]{$q$}
\Text( 30, 30)[lt]{$q$}
\Text( 30,-30)[lb]{$q$}
\end{picture}
\begin{picture}(80,80)(-40,-40)
\SetWidth{0.75}
\ArrowLine(0,0)(-25,-25)
\ArrowLine(25,-25)(0,0)
\ArrowLine(0,0)(-25,25)
\ArrowLine(25,25)(0,0)
\SetColor{Red}
\ArrowLine(0,-10)(-25,-35)
\ArrowLine(25,-35)(0,-10)
\SetColor{Blue}
\ArrowLine(0,10)(-25,35)
\ArrowLine(25,35)(0,10)
\Text(-30,-30)[rb]{$\bar{q}$}
\Text(-30, 30)[rt]{$\bar{q}$}
\Text( 30, 30)[lt]{$\bar{q}$}
\Text( 30,-30)[lb]{$\bar{q}$}
\end{picture}
\begin{picture}(80,80)(-40,-40)
\SetWidth{0.75}
\ArrowLine(-25,-25)(0,0)
\ArrowLine(0,0)(25,-25)
\ArrowLine(0,0)(-25,25)
\ArrowLine(25,25)(0,0)
\SetColor{Red}
\ArrowLine(-25,-35)(0,-10)
\ArrowLine(0,-10)(25,-35)
\SetColor{Blue}
\ArrowLine(0,10)(-25,35)
\ArrowLine(25,35)(0,10)
\Text(-30,-30)[rb]{$q$}
\Text(-30, 30)[rt]{$\bar{q}$}
\Text( 30, 30)[lt]{$\bar{q}$}
\Text( 30,-30)[lb]{$q$}
\end{picture}
\begin{picture}(80,80)(-40,-40)
\SetWidth{0.75}
\ArrowLine(-25,-25)(0,0)
\DashLine(0,0)(25,-25){2}
\ArrowLine(0,0)(-25,25)
\DashLine(25,25)(0,0){2}
\SetColor{Blue}
\ArrowLine(-10,0)(-25,15)
\ArrowLine(-25,-15)(-10,0)
\Text(-30,-30)[rb]{$q$}
\Text(-30, 30)[rt]{$\bar{q}$}
\Text( 30, 30)[lt]{$X^\prime$}
\Text( 30,-30)[lb]{$X$}
\end{picture}
\begin{picture}(80,80)(-40,-40)
\SetWidth{0.75}
\ArrowLine(-25,-25)(0,0)
\ArrowLine(0,0)(-25,25)
\DashLine(0,0)(25,25){2}
\Gluon(0,0)(25,-25){4}{4.5}
\SetColor{Red}
\ArrowLine(-25,-35)(0,-10)
\ArrowLine(0,-10)(25,-35)
\SetColor{Blue}
\ArrowLine(25,-15)(-25,35)
\Text(-30,-30)[rb]{$q$}
\Text(-30, 30)[rt]{$\bar{q}$}
\Text( 30, 30)[lt]{$X$}
\Text( 30,-30)[lb]{$g$}
\end{picture}
\begin{picture}(80,80)(-40,-40)
\SetWidth{0.75}
\Gluon(-25,-25)(0,0){4}{4.5}
\DashLine(0,0)(25,-25){2}
\ArrowLine(-25,25)(0,0)
\ArrowLine(0,0)(25,25)
\SetColor{Red}
\ArrowLine(-25,-35)(0,-10)
\ArrowLine(0,-10)(25,15)
\SetColor{Blue}
\ArrowLine(-10,0)(-25,-15)
\ArrowLine(-25,15)(-10,0)
\Text(-30,-30)[rb]{$g$}
\Text(-30, 30)[rt]{$q$}
\Text( 30, 30)[lt]{$q$}
\Text( 30,-30)[lb]{$X$}
\end{picture}
\begin{picture}(80,80)(-40,-40)
\SetWidth{0.75}
\Gluon(-25,-25)(0,0){4}{4.5}
\DashLine(0,0)(25,-25){2}
\ArrowLine(0,0)(-25,25)
\ArrowLine(25,25)(0,0)
\SetColor{Red}
\ArrowLine(0,-10)(-25,-35)
\ArrowLine(25,15)(0,-10)
\SetColor{Blue}
\ArrowLine(-25,-15)(-10,0)
\ArrowLine(-10,0)(-25,15)
\Text(-30,-30)[rb]{$g$}
\Text(-30, 30)[rt]{$\bar{q}$}
\Text( 30, 30)[lt]{$\bar{q}$}
\Text( 30,-30)[lb]{$X$}
\end{picture}
\begin{picture}(80,80)(-40,-40)
\SetWidth{0.75}
\Gluon(-25,-25)(0,0){4}{4.5}
\Gluon(0,0)(25,-25){4}{4.5}
\ArrowLine(-25,25)(0,0)
\ArrowLine(0,0)(25,25)
\SetColor{Red}
\ArrowLine(-25,-35)(0,-10)
\ArrowLine(0,-10)(25,-35)
\SetColor{Blue}
\ArrowLine(-25,37)(0,12)
\ArrowLine(0,12)(25,37)
\SetColor{Green}
\ArrowLine(0,10)(-25,-15)
\ArrowLine(25,-15)(0,10)
\Text(-30,-30)[rb]{$g$}
\Text(-30, 30)[rt]{$q$}
\Text( 30, 30)[lt]{$q$}
\Text( 30,-30)[lb]{$g$}
\end{picture}
\begin{picture}(80,80)(-40,-40)
\SetWidth{0.75}
\Gluon(-25,-25)(0,0){4}{4.5}
\Gluon(0,0)(25,-25){4}{4.5}
\ArrowLine(0,0)(-25,25)
\ArrowLine(25,25)(0,0)
\SetColor{Red}
\ArrowLine(-25,-35)(0,-10)
\ArrowLine(0,-10)(25,-35)
\SetColor{Blue}
\ArrowLine(0,12)(-25,37)
\ArrowLine(25,37)(0,12)
\SetColor{Green}
\ArrowLine(0,10)(-25,-15)
\ArrowLine(25,-15)(0,10)
\Text(-30,-30)[rb]{$g$}
\Text(-30, 30)[rt]{$\bar{q}$}
\Text( 30, 30)[lt]{$\bar{q}$}
\Text( 30,-30)[lb]{$g$}
\end{picture}
\begin{picture}(80,80)(-40,-40)
\SetWidth{0.75}
\ArrowLine(0,0)(-25,25)
\ArrowLine(-25,-25)(0,0)
\Gluon(0,0)(25,-25){4}{4.5}
\Gluon(0,0)(25,25){4}{4.5}
\SetColor{Red}
\ArrowLine(10,0)(25,15)
\ArrowLine(25,-15)(10,0)
\SetColor{Green}
\ArrowLine(-10,0)(25,-35)
\ArrowLine(25,35)(-10,0)
\SetColor{Blue}
\ArrowLine(-12,0)(-27,15)
\ArrowLine(-27,-15)(-12,0)
\Text(-30,-30)[rb]{$q$}
\Text(-30, 30)[rt]{$\bar{q}$}
\Text( 30, 30)[lt]{$g$}
\Text( 30,-30)[lb]{$g$}
\end{picture}
\begin{picture}(80,80)(-40,-40)
\SetWidth{0.75}
\Gluon(-25,-25)(0,0){4}{4.5}
\ArrowLine(0,0)(25,-25)
\Gluon(-25,25)(0,0){4}{4.5}
\ArrowLine(25,25)(0,0)
\SetColor{Red}
\ArrowLine(27,15)(12,0)
\ArrowLine(12,0)(27,-15)
\SetColor{Blue}
\ArrowLine(-25,15)(-10,0)
\ArrowLine(-10,0)(-25,-15)
\SetColor{Green}
\ArrowLine(-25,-35)(10,0)
\ArrowLine(10,0)(-25,35)
\Text(-30,-30)[rb]{$g$}
\Text(-30, 30)[rt]{$g$}
\Text( 30, 30)[lt]{$\bar{q}$}
\Text( 30,-30)[lb]{$q$}
\end{picture}
\begin{picture}(80,80)(-40,-40)
\SetWidth{0.75}
\Gluon(-25,-25)(0,0){4}{4.5}
\DashLine(0,0)(25,-25){2}
\Gluon(-25,25)(0,0){4}{4.5}
\DashLine(0,0)(25,25){2}
\SetColor{Red}
\ArrowLine(-25,-35)(10,0)
\ArrowLine(10,0)(-25,35)
\SetColor{Blue}
\ArrowLine(-25,15)(-10,0)
\ArrowLine(-10,0)(-25,-15)
\Text(-30,-30)[rb]{$g$}
\Text(-30, 30)[rt]{$g$}
\Text( 30, 30)[lt]{$X^\prime$}
\Text( 30,-30)[lb]{$X$}
\end{picture}
\begin{picture}(80,80)(-40,-40)
\SetWidth{0.75}
\Gluon(-25,-25)(0,0){4}{4.5}
\Gluon(0,0)(25,-25){4}{4.5}
\Gluon(-25,25)(0,0){4}{4.5}
\DashLine(0,0)(25,25){2}
\SetColor{Red}
\ArrowLine(-25,-35)(0,-10)
\ArrowLine(0,-10)(25,-35)
\SetColor{Blue}
\ArrowLine(-25,15)(-10,0)
\ArrowLine(-10,0)(-25,-15)
\SetColor{Green}
\ArrowLine(25,-15)(-25,35)
\Text(-30,-30)[rb]{$g$}
\Text(-30, 30)[rt]{$g$}
\Text( 30, 30)[lt]{$X$}
\Text( 30,-30)[lb]{$g$}
\end{picture}
\begin{picture}(80,80)(-40,-40)
\SetWidth{0.75}
\Gluon(-25,-25)(0,0){4}{4.5}
\Gluon(0,0)(25,-25){4}{4.5}
\Gluon(-25,25)(0,0){4}{4.5}
\Gluon(0,0)(25,25){4}{4.5}
\SetColor{Red}
\ArrowLine(-25,-35)(10,0)
\ArrowLine(10,0)(-25,35)
\SetColor{Blue}
\ArrowLine(-27,15)(-12,0)
\ArrowLine(-12,0)(-27,-15)
\SetColor{Green}
\ArrowLine(27,-15)(12,0)
\ArrowLine(12,0)(27,15)
\SetColor{Magenta}
\ArrowLine(-10,0)(25,-35)
\ArrowLine(25,35)(-10,0)
\Text(-30,-30)[rb]{$g$}
\Text(-30, 30)[rt]{$g$}
\Text( 30, 30)[lt]{$g$}
\Text( 30,-30)[lb]{$g$}
\end{picture}
\end{center}
\caption{\label{fig5}Colour flow diagrams for most the quantum black
hole decays.
The symbol $X$ represents a colourless particle.}
\end{figure}
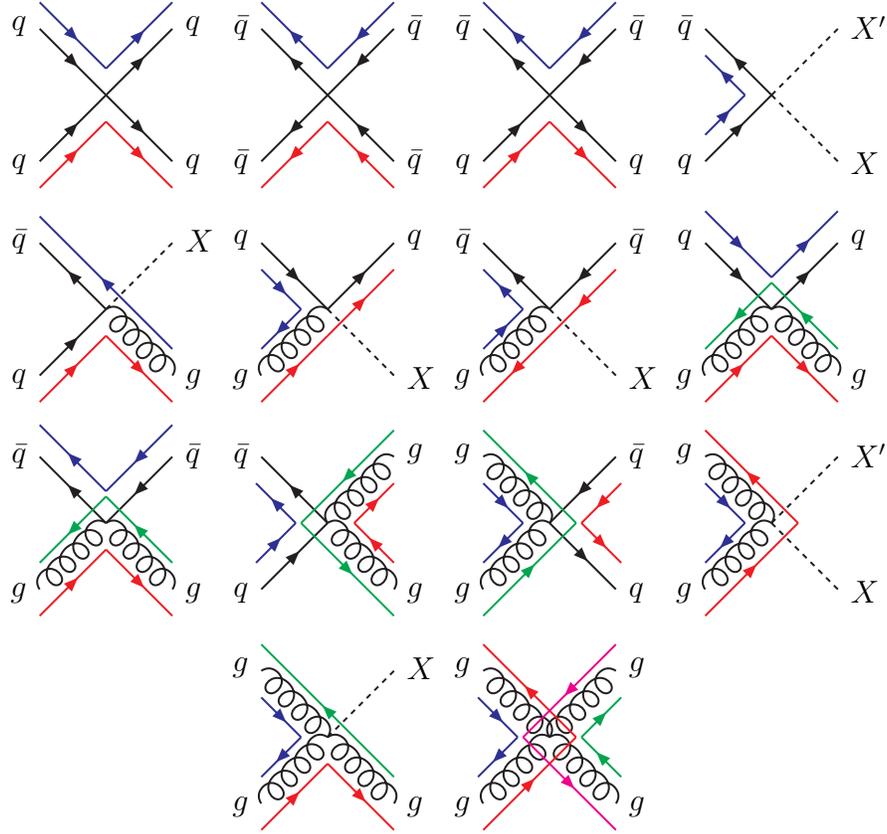

\noindent
\texttt{flow} sets the mothers and colour connections for each final-state
particle. 
In most cases, the colour flow is straightforward.
Figure~\ref{fig5} shows the set of colour flow diagrams for most of the
quantum black hole decay processes.

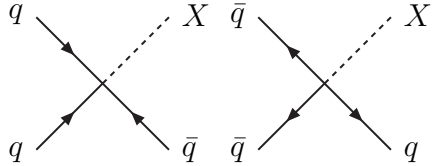
\begin{figure}[htb]
\begin{center}
\begin{picture}(80,80)(-40,-40)
\SetWidth{0.75}
\ArrowLine(-25,-25)(0,0)
\ArrowLine(-25,25)(0,0)
\ArrowLine(25,-25)(0,0)
\DashLine(0,0)(25,25){2}
\Text(-30,-30)[rb]{$q$}
\Text(-30, 30)[rt]{$q$}
\Text( 30, 30)[lt]{$X$}
\Text( 30,-30)[lb]{$\bar{q}$}
\end{picture}
\begin{picture}(80,80)(-40,-40)
\SetWidth{0.75}
\ArrowLine(0,0)(-25,-25)
\ArrowLine(0,0)(-25,25)
\ArrowLine(0,0)(25,-25)
\DashLine(0,0)(25,25){2}
\Text(-30,-30)[rb]{$\bar{q}$}
\Text(-30, 30)[rt]{$\bar{q}$}
\Text( 30, 30)[lt]{$X$}
\Text( 30,-30)[lb]{$q$}
\end{picture}
\end{center}
\caption{\label{fig6}Diagrams for which a nonstandard colour flow
technique has been used.
The symbol $X$ represents a colourless particle.} 
\end{figure}

Figure~\ref{fig6} shows the two diagrams in which the colour flow can
not be implemented naturally, and we have used a non-standard procedure.   
The technique is as follows.
We replace one of the incoming quarks or antiquarks by an antiquark or
quark, respectively, of the same flavour.
The flavour is changed just before the particle is added to the event
record.  
Thus the particle kinematics and weight of the event are unaffected. 
Next, an additional outgoing quark or antiquark is added so that
electric charge is conserved. 
The colour connections can then be made between the coloured particles.
In this way we can conserve electric charge and allow colour to flow.
The additional outgoing particle belongs to the first quark generation
and is given zero momentum-energy.
The extra zero momentum-energy particle adds little to the event.
Typically a colour string will form between it and a beam-remnant
diquark.
String fragmentation will give a few extra baryons and mesons.
Because of the vanishing momentum-energy, the diquark will dominate the
kinematics and typically the extra particles will travel down the beam
pipe undetected.   
Such an approach has been used in previous
generators~\cite{Koch05,Gingrich07a}. 

PYTHIA~8 provides a junction class to deal with baryon violating
decays. 
Currently it handles only the case of one initial state particle.
The facility is not readily available to the users via the Les Houches
Accord interface. 

\bigskip
\noindent
void QuantumBlackHole::deckin(int *idn,dLHAup* lha,double *p1,double *p2)

\begin{tabular}{ll}
Purpose:          & Calculate kinematics of decay.\\
Functions called: & Decay::twobody.\\
Return:           & four-vectors of black hole and final-state particles.\\
\end{tabular}
\bigskip

\noindent
\texttt{deckin} calculates kinematics of the decay.

The following four member functions are private. 

\bigskip
\noindent
void QuantumBlackHole::xpart(double\& xmin,double\& Q,double\& sighat)

\begin{tabular}{ll}
Purpose:          & Calculate parton-parton cross section (pb).\\
Functions called: & none.\\
Return:           & xmin, Q, sighat.\\
\end{tabular}
\bigskip

\noindent
\texttt{xpart} generates a random $x_\mathrm{min}$ and uses this to
calculate the parton-parton differential cross section. 
It also returns $Q$ which is only available after the gravitational
radius and mass are calculated.

\bigskip
\noindent
void QuantumBlackHole::pdf(const double* xx,const double Q)

\begin{tabular}{ll}
Purpose:          & Get parton distributions via PYTHIA \texttt{PDF} class.\\
Functions called: & PDF::xf.\\
Return:           & fills disf[2][13] array.
\end{tabular}
\bigskip

\noindent
\texttt{pdf} gets the parton distribution functions via the PYTHIA PDF class
and stores them as a private data member of the class.
The PDFs are available to this function via a stored private pointer in
the class.

\bigskip
\noindent
void QuantumBlackHole::xsec(const double sighat,double\& dxsec)

\begin{tabular}{ll}
Purpose:          & Calculate the proton-proton cross section (pb).\\
Functions called: & disf[2][13].\\
Return:           & Differential cross section in pb\\
\end{tabular}
\bigskip

\noindent
\texttt{xsec} uses the parton cross section and an array of PDFs to
calculate the proton-proton cross section.

\bigskip
\noindent
void QuantumBlackHole::sample(const double sighat,const double dxsec,
int* idn)

\begin{tabular}{ll}
Purpose:          & Sample cross section to select beam partons.\\
Functions called: & disf[2][13].\\
Return:           & weight, beam parton IDs.\\
\end{tabular}
\bigskip

\noindent
\texttt{sample} samples the cross section to determine the beam partons
and returns a weight.

\subsection{Utility Classes \texttt{Random} and \texttt{Decay}}

The totally static \texttt{Random} class takes care of all the random
distributions and random selection utilities, such as selecting a
random charged lepton, for example. 
The class is driven by the uniform random number generator used by
PYTHIA.
Thus PYTHIA may be used to initialize and manage the seeds of the
\texttt{Random} class.

The \texttt{Decay} class handles the two-particle phase space decay.
It has one public member to perform the decays.
A set of private member functions perform all the required Lorentz
transformations. 

\section{Installation and Availability}

The QBH generator is written entirely in C++.
It interfaces only to PYTHIA~8.
If a lower-bound on the cross section is being calculated by using the
trapped surface option, then the appropriate file is read in.
These small text files must exist in the same directory as as the
executable image.
If the LHAPDF parton densities are required, the LHAPDF library must be
linked.
The generator has been tested with PYTHIA version 8.150 and LHAPDF
version 5.8.0.
Version 1.02 and earlier of the generator should be used with version
8.130 of PYTHIA.
Both Linux and Darwin-gcc4 architectures with the slc4\_ia32\_gcc34 platform
have been tested.

The source code can be obtain from the HepForge site
http://projects.hepforge.org/qbh/. 

\section{Performance}

\begin{figure}[t]
\begin{center}
\includegraphics[width=13.5cm]{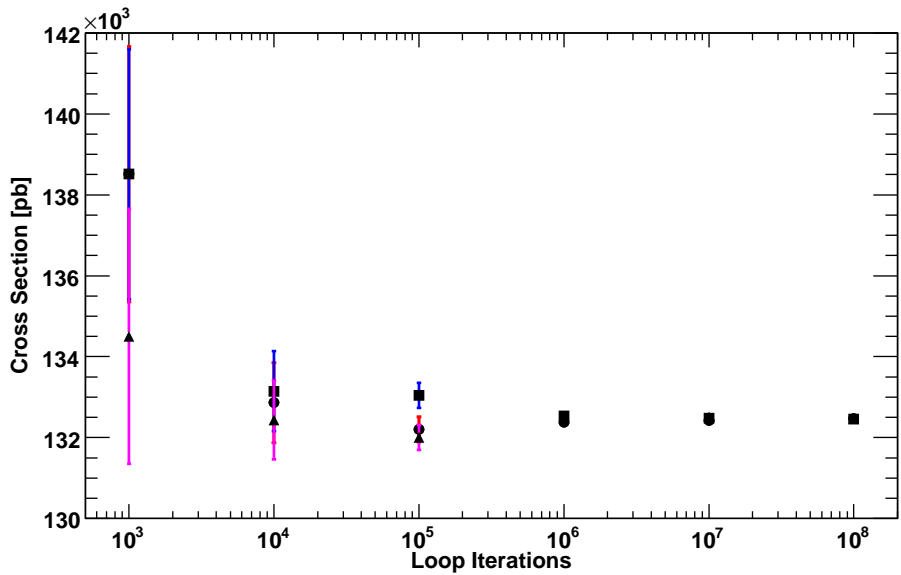}
\caption{Total cross section versus number of loop iterations used in the
cross section calculation.}   
\label{fig10}
\end{center}
\end{figure}

The performance of the event generator has been measured.
Figure~\ref{fig10} shows the cross section versus number of loop
iterations use to calculate the cross section for three different runs
at each value of the number loop iterations.
The spread in the three cross section values at each number of
loop iterations is consistent with the cross section statistical error
bars. 
The cross section calculation is stable and converges nicely.
For a statistical accuracy of 0.2\%, $10^5$ iterations are required,
while for an accuracy of 0.07\%, $10^6$ iterations are required.
The totally inelastic cross sections have been used for the
performance studies.
One might expect the trapped surface cross sections to take a little
longer. 

As a further check of the cross section calculation, we compared the
cross section calculated by QBH using $10^5$ loop iterations with that
calculated by PYTHIA for $10^5$ generated events.  
The percentage difference was less than 0.5\%.

\begin{figure}[htb]
\begin{center}
\includegraphics[width=13.5cm]{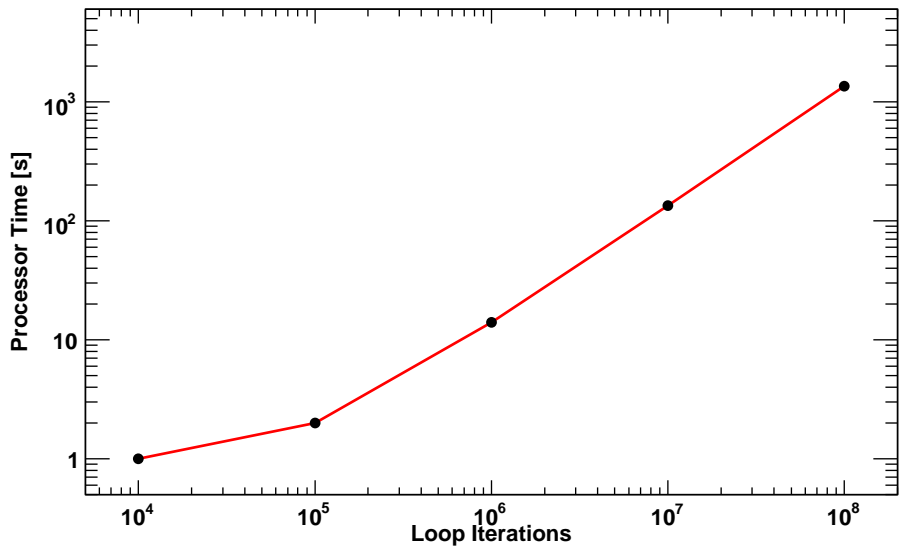}
\caption{Processor time versus number of loop iterations in the total cross
section calculation.
The timing resolutions is 1~s and the processor time axis should be
considered as a relative timing.}
\label{fig11}
\end{center}
\end{figure}

Figure~\ref{fig11} shows the processor time versus number of loop
iterations in the cross section calculation.
To obtain a 0.2\% accuracy requires 2~s, while for a 0.07\% accuracy 14~s
is needed.
However the processor used was arbitrary so the time axis should be
viewed as a relative time.
By default, the generator used $10^5$ loop iterations to calculate the
cross section.

\begin{figure}[htb]
\begin{center}
\includegraphics[width=13.5cm]{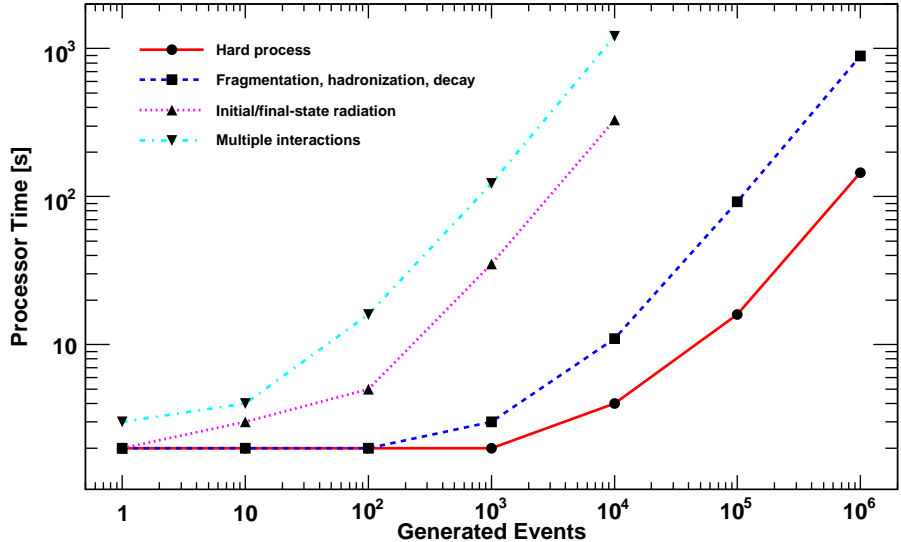}
\caption{Processor time versus number of events generated.
The timing resolutions is 1~s.
The processing tasks are cumulative.  
For example, multiple interactions also includes all the other
processing.} 
\label{fig12}
\end{center}
\end{figure}

Figure~\ref{fig12} shows the processor time versus number of generated
events. 
To decouple the generator from PYTHIA~8, the ``hard process'' times have
fragmentation, hadronization, particle decays, initial- and final-state
radiation, and multiple interactions switched off.
About $6\times 10^3$ QBH events can be generated per second.
If fragmentation, hadronization, and decays are turned on, about
$1\times 10^3$ events can be generated per second.
Turning on initial- and final-state radiation allows 30 events per
second to be generated.
Finally, turning on multiple interactions (everything) allows 8 events 
per second to be generated.

\section*{Acknowledgments}

This work was supported in part by the Natural Sciences and Engineering
Research Council of Canada.





\bibliographystyle{elsarticle-num}
\bibliography{gingrich}

\begin{thebibliography}{10}
\expandafter\ifx\csname url\endcsname\relax
  \def\url#1{\texttt{#1}}\fi
\expandafter\ifx\csname urlprefix\endcsname\relax\def\urlprefix{URL }\fi
\expandafter\ifx\csname href\endcsname\relax
  \def\href#1#2{#2} \def\path#1{#1}\fi

\bibitem{Gingrich09b}
D.~M. Gingrich, Quantum black holes with charge, color and spin at the {LHC},
  J. Phys. G: Nucl. Part. Phys. 37 (2010) 105008.
\newblock \href {http://arxiv.org/abs/arXiv:0912.0826v4 [hep-ph]}
  {\path{arXiv:arXiv:0912.0826v4 [hep-ph]}}.

\bibitem{pythia8}
T.~Sj{\"o}strand, S.~Mrenna, P.~Skands, A brief introduction to {PYTHIA} 8.1,
  Comput. Phys. Commun. 178 (2008) 852--867.
\newblock \href {http://arxiv.org/abs/arXiv:0710.3820v1}
  {\path{arXiv:arXiv:0710.3820v1}}.

\bibitem{pythia64}
T.~Sj{\"o}strand, S.~Mrenna, P.~Skands, {PYTHIA} 6.4 physics and manual, J.
  High Energy Phys. 0605 (2006) 026.
\newblock \href {http://arxiv.org/abs/arXiv:hep-ph/0603175v2}
  {\path{arXiv:arXiv:hep-ph/0603175v2}}.

\bibitem{LHA}
{E. Boos \textit{et al.}}, Generic user interface for event generators,
  arXiv:hep-ph/0109068v1 (2001).

\bibitem{Dimopoulos01}
S.~Dimopoulos, G.~Landsberg, Black holes at the {L}arge {H}adron {C}ollider,
  Phys.\ Rev.\ Lett. 87 (2001) 161602.
\newblock \href {http://arxiv.org/abs/arXiv:hep-ph/0106295v1}
  {\path{arXiv:arXiv:hep-ph/0106295v1}}.

\bibitem{Tanaka}
J.~Tanaka, T.~Yamamura, S.~Asai, J.~Kanzaki, Study of black holes with the
  {ATLAS} detector at the {LHC}, Eur.\ Phys.\ J. C 41 (2005) s02 19--33.
\newblock \href {http://arxiv.org/abs/arXiv:hep-ph/0411095}
  {\path{arXiv:arXiv:hep-ph/0411095}}.

\bibitem{Harris03a}
C.~M. Harris, P.~Richardson, B.~R. Webber, {CHARYBDIS}: {A} black hole event
  generator, J. High Energy Phys. 08 (2003) 033.
\newblock \href {http://arxiv.org/abs/arXiv:hep-ph/0307305}
  {\path{arXiv:arXiv:hep-ph/0307305}}.

\bibitem{Cavaglia06}
M.~Cavagli{\`a}, R.~Godang, L.~Cremaldi, D.~Summers, {C}atfish: A {M}onte
  {C}arlo simulator for black holes at the {LHC}, Comput. Phys. Commun. 177
  (2007) 506--517.
\newblock \href {http://arxiv.org/abs/arXiv:hep-ph/0609001v2}
  {\path{arXiv:arXiv:hep-ph/0609001v2}}.

\bibitem{Gingrich06b}
D.~M. Gingrich, Comparison of black hole generators for the {LHC},
  arXiv:hep-ph/0610219v2 (2006).

\bibitem{Frost09}
J.~A. Frost, J.~R. Gaunt, M.~O.~P. Sampaio, M.~Casals, S.~R. Dolan, M.~A.
  Parker, B.~R. Webber, Phenomenology of production and decay of spinning
  extra-dimensional black holes at hardon colliders, J. High Energy Physics 10
  (2009) 014.
\newblock \href {http://arxiv.org/abs/arXiv:0904.0979v4 [hep-ph]}
  {\path{arXiv:arXiv:0904.0979v4 [hep-ph]}}.

\bibitem{Dai09}
D.-C. Dai, C.~Issever, E.~Rizvi, G.~Starkman, D.~Stojkovic, J.~Tseng, Manual of
  {B}lack{M}ax, a black-hole event generator with rotation, recoil, split
  branes, and brane tension, arXiv::0902.3577v1 [hep-ph] (2009).

\bibitem{Meade08}
P.~Meade, L.~Randall, Black holes and quantum gravity at the {LHC}, J. High
  Energy Physics 05 (2008) 003.
\newblock \href {http://arxiv.org/abs/arXiv:0708.3017v1 [hep-ph]}
  {\path{arXiv:arXiv:0708.3017v1 [hep-ph]}}.

\bibitem{LHALEF}
{J. Alwall \textit{et al.}}, A standard format for {L}es {H}ouches event files,
  Comput. Phys. Commun. 176 (2007) 300--304.
\newblock \href {http://arxiv.org/abs/arXiv:hep-ph/0609017v1}
  {\path{arXiv:arXiv:hep-ph/0609017v1}}.

\bibitem{Yoshino05a}
H.~Yoshino, V.~S. Rychkov, Improved analysis of black hole formation in
  high-energy particle collisions, Phys.\ Rev.\ D 71 (2005) 104028.
\newblock \href {http://arxiv.org/abs/arXiv:hep-th/0503171v2}
  {\path{arXiv:arXiv:hep-th/0503171v2}}.

\bibitem{Gingrich06a}
D.~M. Gingrich, Black hole cross-section at the {LHC}, Int. J. Mod. Phys. A 21
  (2006) 6653--6676.
\newblock \href {http://arxiv.org/abs/arXiv:hep-ph/0609055v2}
  {\path{arXiv:arXiv:hep-ph/0609055v2}}.

\bibitem{LHAPDF}
{LHAPDF the Les Houches Accord PDF Interface},
  http://hepforge.cedar.ac.uk/lhapdf/.

\bibitem{Koch05}
B.~Koch, M.~Bleicher, S.~Hossenfelder, Black hole remnants at the {LHC}, J.
  High Energy Phys. 0510 (2005) 053.
\newblock \href {http://arxiv.org/abs/arXiv:hep-ph/0507138v2}
  {\path{arXiv:arXiv:hep-ph/0507138v2}}.

\bibitem{Gingrich07a}
D.~M. Gingrich, Missing energy in black hole production and decay at the
  {L}arge {H}adron {C}ollider, J. High Energy Phys. 11 (2007) 064.
\newblock \href {http://arxiv.org/abs/arXiv:0706.0623v2 [hep-ph]}
  {\path{arXiv:arXiv:0706.0623v2 [hep-ph]}}.

\end{thebibliography}







\end{document}